\documentclass[12pt]{article}
\usepackage{latexsym}
\usepackage{amsmath}
\usepackage{amsfonts}
\usepackage{amssymb}
\def\hybrid{\topmargin -20pt    \oddsidemargin 0pt
        \headheight 0pt \headsep 0pt
        \textwidth 6.25in       
        \textheight 9.5in       
        \marginparwidth .875in
        \parskip 5pt plus 1pt   \jot = 1.5ex}

\hybrid

\newcommand{\beq}{\begin{equation}}
\newcommand{\eeq}{\end{equation}}
\newcommand{\bi}{\begin{itemize}}
\newcommand{\ei}{\end{itemize}}
\newcommand{\bea}{\begin{eqnarray}}
\newcommand{\eea}{\end{eqnarray}}
\newcommand{\ba}{\begin{array}}
\newcommand{\ea}{\end{array}}
\newcommand{\bt}{\begin{tabular}}
\newcommand{\et}{\end{tabular}}
\newcommand{\bc}{\begin{center}}
\newcommand{\ec}{\end{center}}

\def\theequation{\arabic{section}.\arabic{equation}}

\newcommand{\ket}[1]{|#1\rangle}

\newcommand{\vev}[1]{\langle#1\rangle}

\newcommand{\im}{\text{Im}}
\newcommand{\ir}[1]{\hat#1}

\begin{document}

\begin{titlepage}
\begin{center}

\hfill hep-th/0403128\\
\hfill LPTENS 04/11\\

\vskip 1.5cm
{\Large \bf Spontaneous $N=2\rightarrow N=1$ Supergravity Breaking \\
in Three Dimensions}

\vskip 1.5cm

{\bf Olaf Hohm$^{\rm a}$ and Jan Louis$^{\rm a,b}$} \\

\vskip 20pt

${}^{\rm a}${\em II. Institut f\"ur Theoretische Physik\\ 
Universit\"at Hamburg\\
Luruper Chaussee 149\\
D-22716 Hamburg, Germany}\\

\vskip 15pt

${}^{\rm b}${\em Laboratoire de Physique Th\'eorique de l'Ecole Normale 
Sup\'erieure\\
24 rue Lhomond, \\
75231 Paris Cedex, France}\\

\vskip 10pt

{email: {\tt olaf.hohm@desy.de, jan.louis@desy.de}} \\

\vskip 0.8cm

\end{center}

\vskip 2cm

\begin{center} {\bf ABSTRACT} \end{center}

\noindent

 We study models of spontaneous 
 $N=2\to N=1$ supergravity breaking in three space-time dimensions and
 discuss the topological Higgs- and super-Higgs mechanism which 
 generates the masses for the spin-$3/2$ gravitino multiplet.
 The resulting $N=1$ spectrum and its effective action is
 analysed.

\vfill

March 2004

\end{titlepage}

\section{Introduction}

\setcounter{equation}{0}
Partial supersymmetry breaking is of interest in its own right
but also within string theory. The 
early no-go theorems  stated that partial supersymmetry 
breaking is impossible and either all supercharges are 
preserved or all supercharges are broken \cite{CGP}.
These theorems were later on 
modified in global supersymmetry \cite{APT,BG},
supergravity \cite{FGP} 
and in string theory \cite{HP,KK,TV,PM1,CKLT}.
However, a satisfactory and conceptual understanding of partial
supersymmetry breaking is still lacking.

In supergravity 
the main focus so far was on spontaneous $N=2 \to N=1$ 
in four space-time dimensions ($D=4$) \cite{FGP,FGPT,louis1,Ferrara}.
In this paper we concentrate on the somewhat simpler situation of 
$N=2 \to N=1$ breaking in $D=3$. 
These $N=2, D=3$ theories are closely related to dimensionally reduced
$N=1$ theories in $D=4$ as they also feature four supercharges.
Partial breaking to $N=1$ (in $D=3$) breaks two of the four supercharges 
leaving an $N=1$ in $D=3$ intact.
In the context of global supersymmetry this breaking has also 
been discussed in \cite{IK}.

The reason to concentrate on $D=3$ vacua is on the one hand 
their simplicity. On the other hand they also arise
in the context of string theory by compactifying M-theory 
on Calabi-Yau fourfolds  with 
non-trivial background fluxes \cite{BB,GVW,DSR,haack,GSS,BBHL,BHS}.
The background fluxes appear in the low energy effective action 
as gauge or mass parameters and generically
turn an ordinary supergravity into a gauged
or massive supergravity. 
The fluxes induce a potential for the scalar fields 
which only under certain conditions preserves some of the supercharges.
A class of  models where Peccei-Quinn isometries are gauged 
and $N=2\to N=1$  breaking occurs was recently found in ref.\ 
\cite{BHS} and in our analysis we closely follow this work.

The purpose of this paper is to further analyse
the mechanism of $N=2\to N=1$ breaking in $D=3$.
In particular we study the details of the Higgs and super-Higgs effect
which gives masses to  the gauge bosons and the gravitino.
We find that contrary to $D=4$ a topological Higgs mechanism 
is at work in $D=3$ where no Goldstone boson is eaten \cite{schonfeld,deser1}. 
The presence of a Chern-Simons term in addition to the standard
Yang-Mills terms renders a gauge field massive without
adding any physical degree of freedom to the gauge field. 
Alternatively in a `dual' version of this mechanism
the gauge theory has no standard
Yang-Mills term but instead a Chern-Simons term and a conventional
mass term \cite{jackiw,townsend} or in other words a Chern-Simons
term and a coupling of the gauge field to a Goldstone boson \cite{deser2}.

More specifically 
the paper is organized as follows.
In section 2 we recall the necessary facts  
about gauged $N=2$ supergravity in $D=3$ following 
refs.\ \cite{dWNT,dWHS,dWNS} with particular emphasis
on gauged Peccei-Quinn isometries \cite{BHS}. 
In section 3 we review 
the class of models considered in \cite{BHS}
where spontaneous 
$N=2\rightarrow N=1$ supergravity breaking occurs.
 The super-Higgs mechanism and the
reorganization in $N=1$ supermultiplets are discussed in section 4. 
Section 5 contains our conclusions and our spinor conventions are given in 
appendix A. In appendix B we review the construction of the massive 
supermultiplets for arbitrary $N$ in $D=3$ which are necessary for the 
analysis of the possible multiplets resulting from partial 
supergravity breaking.

\section{Gauged $N=2$ Supergravity in $D=3$}
\setcounter{equation}{0}
\subsection{Generalities}

Let us start by  first recalling a few facts about 
$N=2$ supergravity in $D=3$ \cite{dWNT,dWHS}. 
This theory is closely related to a dimensionally reduced $N=1$ supergravity
in $D=4$ and as a consequences similar
multiplets and geometrical structures appear.

The gravity multiplet $(g_{\mu\nu},\psi_{\mu}^I)$, 
$\mu,\nu=0,1,2,\, I=1,2$, contains 
besides the space-time metric $g_{\mu\nu}$ two Majorana gravitini 
$\psi_{\mu}^{I}$. (See appendix A for our spinor conventions and appendix~B
for a summary of the supermultiplets.)
In contrast to higher-dimensional theories these fields carry no 
propagating degrees of freedom. 
In order to get dynamically non-trivial theories
one can add $d$ $N=2$ scalar multiplets $(\phi^i,\chi^i_1,\chi^i_2)$, 
$i=1,...,d$, consisting of $d$ complex scalar
fields $\phi^i$ and $2d$ real Majorana fermions $\chi^i_1,\chi^i_2$.
The bosonic Lagrangian is  given by
\begin{equation}\label{sigma}
  \mathcal{L}_{\text{B}}=\frac{1}{2}\, e R 
- e g_{i\bar{j}}(\phi,\bar{\phi})\,
  \partial_{\mu}\phi^i \partial^{\mu}\bar{\phi}^{\bar{j}} + eV \ ,
 \end{equation}
where $g_{i\bar{j}}$ is a metric on a K\"ahler manifold
which can be 
expressed in terms of a K\"ahler potential as
$g_{i\bar{j}}=\partial_i\partial_{\bar{j}}K$.
$V$ is determined in terms of a holomorphic superpotential 
$W(\phi)$ and its K\"ahler 
covariant derivative
$D_iW=\partial_iW + (\partial_iK)W$ via\footnote{
Following \cite{dWHS}  we have included
 a factor $g^2$ in the potential for later convenience.
In a more standard notation it would be reabsorbed in the definition
of $W$.}
\begin{equation}\label{pot1}
  V=\frac{g^2}{4}\,e^K(g^{i\bar{j}}D_i W D_{\bar{j}}
  \overline{W}- 4 |W|^2)\ .
 \end{equation}

It is possible to couple gauge fields
to this theory by gauging isometries of the metric $g_{i\bar{j}}$.
Such isometries are generated by $n$ holomorphic Killing vector fields 
$X^i_A$ satisfying 
 \begin{equation}\label{Killing}
\partial_{\bar{j}}X^i_A=0\ , \qquad 
\nabla_i X_{\bar{j} A}+\nabla_{\bar{j}}X_{iA}=0\ ,\qquad  A= 1, \ldots, n\ .
 \end{equation}
These Killing equations determine  $X^{\bar{j}}_A$
in terms of Killing prepotentials (or momentum maps) 
$\mathcal{P}_A$ via
\begin{equation}\label{impuls}
g_{k\bar{j}}X^{\bar{j}}_A = -2i
  \partial_k\mathcal{P}_A\ .
 \end{equation} 
Under the isometries  the
scalar fields transform according to 
 \begin{equation}\label{iso}
  \delta\phi^i = \alpha^A X_A^i(\phi)\ ,
 \end{equation}
where $\alpha^A$ are the (local) gauge parameters for 
the $n$ linearly independent isometries.
Gauge invariance is ensured by introducing  
$n$ vector fields 
$A_{\mu A}$ and promoting the ordinary derivative $\partial_{\mu}\phi$
to covariant derivatives 
 \begin{equation}\label{covariant}
  \mathcal{D}_{\mu}\phi^i=
\partial_{\mu}\phi^i+g\Theta^{AB}A_{\mu A} X^{i}_B\ ,
 \end{equation}        
where $g$ is  the gauge coupling  and $\Theta^{AB}$
an arbitrary constant symmetric matrix.

Introducing the covariant derivative alone is not sufficient
but in addition the supersymmetry transformation laws have to be 
modified which in turn requires adding 
Yukawa-type couplings for the fermionic fields and a scalar 
potential. Furthermore, a ``kinetic'' term for the vector fields
has to be supplied. 

The usual procedure would be to promote $A_{\mu A}$ to a supermultiplet
and add a standard supersymmetric Yang-Mills kinetic term to the action.
However, in $D=3$ there is another peculiar option advisable \cite{NS}.
It is necessary to add instead 
 a Chern-Simons term for the gauge fields
 \begin{equation}\label{chern}
  \mathcal{L}_{\text{CS}}=\frac{1}{4}g\Theta^{AB}
  \varepsilon^{\mu\nu\rho}A_{\mu A}F_{B\nu\rho} \ ,
 \end{equation}   
which is topological and therefore introduces 
no propagating gauge degrees of freedom.
Or in other words the gauge fields are auxiliary fields which
nevertheless ensure the gauge invariance of the theory.
Since $A_{\mu A}$ with the interaction (\ref{chern}) and (\ref{covariant})
has no physical degree of freedom it is not necessary
to introduce additional fermionic partners in order to balance
the boson-fermion degeneracy. Indeed, the consistency with
supersymmetry has been established in \cite{dWHS}. 

The scalar potential $V$ given in (\ref{pot1})
has to be modified in gauged supergravities according 
to  \cite{BHS,dWHS}
\begin{equation}\label{pot}
  V=g^2(4g^{i\bar{j}}\partial_i T\partial_{\bar{j}}T
  -4T^2+ \frac{1}{4}\, e^K(g^{i\bar{j}}D_i W D_{\bar{j}}
  \overline{W}-4|W|^2))\ ,
 \end{equation}
where $T=\mathcal{P}_A\Theta^{AB}\mathcal{P}_B$
($\mathcal{P}_A$ is the momentum map defined in 
(\ref{impuls})). 
 $T$ and $W$ are not completely independent
but gauge invariance of $W$ imposes the constraint \cite{dWHS}
  \begin{equation}\label{constraint}
   X_A^iD_iW = 2iW\mathcal{P}_A \ .
  \end{equation}
To summarize, the bosonic part of the Lagrangian for an  
$N=2$ action for gauged supergravity has the generic form
 \begin{equation}\label{boson}
  e^{-1}\mathcal{L}=\frac{1}{2}R-g_{i\bar{j}}\mathcal{D}_{\mu}\phi^i
  \mathcal{D}^{\mu}\bar{\phi}^{\bar{j}}
  -\frac{1}{4}g\Theta^{AB}\varepsilon_{\mu\nu\rho}
  A^{\mu}_AF^{\nu\rho}_B+V \ .
 \end{equation}          

Let us now turn to the fermionic couplings. 
It is more convenient to treat the $2d$ Majorana fermions $\chi^{\ir{j}}$
as real fermions (i.e.\ not assemble them in a complex notation
as we did for the bosons) and label them
by the `real' indices $\ir{i},\ir{j} = 1,...,2d$.
In this notation the K\"ahler metric is denoted by $g_{\ir{i}\ir{j}}$
which when written in complex coordinates ($\ir{i} = (i,\bar i)$)
only has the non-vanishing components $g_{i\bar{j}}=g_{\bar{i}j}$
known from (\ref{sigma}).
In this paper we only need the fermion bilinear terms
which read \cite{dWHS}
\bea\label{yukawa}
  e^{-1}\mathcal{L}_{\text{F}} &=&
-\frac{i}{2}\, \varepsilon^{\mu\nu\rho}
  \bar{\psi}_{\mu}^I\nabla_{\nu}\psi_{\rho}^I
-\frac{1}{2}\, g_{\ir{i}\ir{j}}\bar{\chi}^{\ir{i}}\gamma^{\mu}
  \nabla_{\mu}\chi^{\ir{j}}\\
&&+
\frac{1}{2}gA_1^{IJ}(\phi)\bar{\psi}_{\mu}^I
  \gamma^{\mu\nu}\psi_{\nu}^J + 2gA_{2\ir{j}}^{I}(\phi)\bar{\psi}_{\mu}^I
  \gamma^{\mu}\chi^{\ir{j}}+2igA_{3\ir{i}\ir{j}}(\phi)
  \bar{\chi}^{\ir{i}}\chi^{\ir{j}}\ .\nonumber
 \eea
Furthermore, supersymmetry demands that the tensors $A_1$ and 
$A_2$ satisfy a quadratic identity which
also determines the potential. It reads
\begin{equation}\label{quad2}
  A_1^{IK}A_1^{KJ}-g^{\ir{i}\ir{j}}A_{2\ir{i}}^{I}A_{2\ir{j}}^{J}
  =-g^{-2}V\delta^{IJ}.
 \end{equation}
In addition, at the stationary points of the potential, i.e.
at the points where the first derivatives vanish, the following
identity is valid \cite{dWHS}
 \begin{equation}\label{quad3}
  3A_1^{IJ}A_{2\ir{i}}^{J}+g^{\ir{j}\ir{k}}A_{2\ir{j}}^{I}
  A_{3\ir{k}\ir{i}}=0\ .
 \end{equation}
Explicitly, in our conventions 
the gravitino mass matrix $A_1$ is given by \cite{dWHS}
  \begin{equation}\label{a1}
  A_1^{IJ}=\left(\begin{array}{cc} -2T & 0 \\ 0 & -2T \end{array}\right)
  +e^{K/2}\left(\begin{array}{cc} -\text{Re}W & \text{Im}W \\ 
  \text{Im}W & \text{Re}W \end{array}\right)\ ,
 \end{equation} 
while $A_2$ and $A_3$ read
in complex notation 
 \begin{equation}
  \begin{split}\label{a2}
   A_{2i}^1&=-\frac{1}{2}(\partial_iT + e^{K/2}D_iW), \\ 
   A_{2i}^2&=\frac{i}{2}(\partial_iT-e^{K/2}D_iW),\\
   A_{3ij}&=\frac{1}{4}e^{K/2}D_iD_jW, \\
   A_{3i\bar{j}}&=-D_i\partial_{\bar{j}}T-\frac{1}{2}g_{i\bar{j}}T
   +\partial_i\partial_{\bar{j}}T.\ 
  \end{split}
 \end{equation}

\subsection{Gauged Peccei-Quinn-Isometries}\label{redual}

A special class of isometries are translational 
or Peccei-Quinn (PQ) isometries. They correspond to
constant Killing vectors $X_A^i$ and as we see from (\ref{iso})
they act as (local)  shifts
$\phi\rightarrow \phi + \alpha$ on a subset of the scalar fields. 
Such isometries typically appear in the effective low energy
supergravities arising from string theory when background fluxes
are turned on. For this reason they were  
studied in ref.\ \cite{BHS} and in this section we follow their analysis.

Since the PQ isometries act on a subset of the scalar 
fields it is convenient  to introduce a notation
where the scalar fields $\phi^i$ are split
into
 \begin{equation}\label{coordinates}
  \phi^i=(\phi^a,\phi^A) = (\phi^a, \varphi^A+i\hat{\varphi}^A)\  ,\quad 
a=1,\ldots,d-n,\ A=1,\ldots,n.
 \end{equation}
We define the PQ symmetries to act only on the 
$\hat\varphi^A$ but that they leave the $\phi^a$ and $\varphi^A$
invariant. Using (\ref{iso}) this corresponds to the Killing vectors
 \begin{equation}\label{kill}
  X^i_B=(0,i\delta_B^A)\ .
 \end{equation}
Inserted into (\ref{covariant}) yields
 \begin{equation}\label{covariant2}
  \mathcal{D}_{\mu}\hat{\varphi}^A
  =\partial_{\mu}\hat{\varphi}^A+g\Theta^{AB}A_{\mu B} \ ,
 \end{equation}
while the covariant derivatives of all other scalars
reduce to ordinary partial derivatives.

The PQ-symmetries severely constrain the possible
couplings of the charged scalar fields.
In particular the potential $V$ (and thus $W$ and $T$)
and the K\"ahler potential $K$ cannot be arbitrary.
For simplicity ref.\ \cite{BHS} assumed that
all these couplings are independent of 
$\hat{\varphi}^A$.\footnote{This situation
certainly has a PQ-symmetry but it is not the most general case.}
For the holomorphic superpotential this implies
by the constraint (\ref{constraint}) that
$W$ is only a function of the $\phi^a$, while the K\"ahler potential
obeys $K=K(\phi^a,\bar \phi^{\bar{a}}, \phi^A + \bar\phi^A)$.
This in turn implies
\begin{equation}\label{metric}
   g_{i\bar{j}}=\left(\begin{array}{cc} g_{a\bar{b}} & g_{a\bar{B}} \\ 
   g_{A\bar{b}} & g_{A\bar{B}} \end{array}\right)
   =\left(\begin{array}{cc} g_{a\bar{b}} & g_{aB} \\ 
   g_{A\bar{b}} & g_{AB} \end{array}\right),
 \end{equation}  
where $g_{a\bar{b}}=\partial_a\partial_{\bar{b}}K$,
$g_{aA}=\frac{1}{2}\partial_a\partial_A K$, $g_{AB}=\frac{1}{4}
\partial_A\partial_B K=:\frac{1}{2}G_{AB}$ and $\partial_A$ denotes a real 
derivative.

In $D=3$ a massless propagating vector is 
Hodge-dual to a massless scalar. Under certain
condition this duality can be inverted in that
a massless scalar can be dualized `back' to a vector.
In \cite{BHS} it was shown that this is possible
precisely for the scalars $\hat{\varphi}^A$
that are charged under the PQ-symmetry and that
the duality relation is given by
\begin{equation}\label{dual2}
  F_{A\mu\nu}=-\varepsilon_{\mu\nu\rho}(G_{AB}\mathcal{D}^{\rho}\hat{\varphi}^B
  +2\im [g_{aA}\partial^{\rho}\phi^a])
  +\text{fermions}\ .
 \end{equation}
If one assumes that $G_{AB}$ is 
invertible, (\ref{dual2}) can be used to express
${\mathcal{D}}^{\rho}\hat{\varphi}^B$ in terms of
$F_{A\mu\nu}$ such that the entire Lagrangian
no longer depends on $\hat{\varphi}^B$.
Instead, the vector fields $A_{\mu A}$ obtain 
proper Yang-Mills kinetic terms and become  
propagating degrees of freedom.
In terms of supermultiplets the original 
scalar multiplet is dualized into a vector multiplet
which contains a vector $A_\mu$, two Majorana
gauginos $\chi^1, \chi^2$ and a real scalar
$\varphi$.
In these field variables the bosonic Lagrangian reads
\cite{BHS}
 \begin{equation}\label{wirkung}
  \begin{split}
   e^{-1}\mathcal{L} =&\frac{1}{2}R-\frac{1}{2}G^{AB}\partial_{\mu}M_A
   \partial^{\mu}M_B+\frac{1}{4}G^{AB}F_A^{\mu\nu}F_{B\mu\nu}
   -G_{a\bar{b}}\partial_{\mu}\phi^a\partial^{\mu}\bar{\phi}^{\bar{b}} \\
   &+\varepsilon_{\mu\nu\rho}F_A^{\mu\nu}\text{Im}
   [G^{AB}g_{aB}\partial^{\rho}\phi^a]+\frac{1}{4}g\Theta^{AB}
  \varepsilon_{\mu\nu\rho}A_A^{\mu}F_B^{\nu\rho}+V\ ,
  \end{split}
 \end{equation}   
where in addition the $\varphi^A$ are eliminated
in favor of the 
(real) coordinates 
$M_A:=\frac{1}{2}\partial_A K$.
Furthermore, the K\"ahler metric is redefined
such that \cite{BHS}\footnote{Note that $K$ is still 
the K\"ahler potential for the 
metric $g_{a\bar{b}}$ 
but not for the redefined $G_{a\bar{b}}$.}
 \begin{equation}\label{strangemetric}
   G_{a\bar{b}}=(G^{a\bar{b}})^{-1}:=g_{a\bar{b}}-2g_{aA}G^{AB}g_{\bar{b}B} 
   =(g^{a\bar{b}})^{-1} \ ,\qquad
   G^{a\bar{b}}=g^{a\bar{b}}\ .
 \end{equation}
Finally, using (\ref{impuls}) and (\ref{kill})
the Killing prepotential
$\mathcal{P}_A$ is found to be
\begin{equation}
  \mathcal{P}_A=\frac{1}{4}\partial_A K 
= \frac{1}{2}M_A\ ,
 \end{equation} 
resulting in 
 \begin{equation}\label{T}
  T=\frac{1}{4}M_A\Theta^{AB}M_B\ .
 \end{equation} 
Inserted into (\ref{pot}) one arrives at
 \begin{equation}\label{pot2}
 \begin{split}
   g^{-2}V\ =\ &\frac{1}{2}M_A\Theta^{AC}G_{CD}\Theta^{DB}M_B
   -\frac{1}{4}(M_A\Theta^{AB}M_B)^2 \\
   &+\frac{1}{4}e^K G^{a\overline{b}}D_a W D_{\overline{b}}\overline{W} 
   -e^K(1-\frac{1}{2}M_A G^{AB}M_B)|W|^2\ ,
  \end{split}
 \end{equation}
where the K\"ahler-covariant derivative is given by 
$D_aW=\partial_aW +(\partial_aK)W$ and $\partial_aK$ depends in general
also on $M_A$.

\section{Conditions for Partial
$N=2\rightarrow N=1$ Breaking}\label{breaking}
\setcounter{equation}{0}
Let us now determine the necessary condition
such that the $N=2$ theory can show partial supersymmetry
breaking to $N=1$ \cite{BHS}. We start from a 
generic $N=2$
supergravity spectrum with 
a gravitational multiplet, $n$ vector multiplets
and $d-n$ scalar multiplets
 \begin{equation}\label{spectrum}
[g_{\mu\nu},\psi^1_{\mu},\psi^2_{\mu}]
  \oplus[A_{A\mu},\chi_A^1,\chi_A^2,M_A]\oplus
  [\phi^a,\chi^a_1,\chi^a_2]\ .
 \end{equation} 

The condition for unbroken supersymmetry
is the vanishing of the fermionic supersymmetry variations in the 
(Lorentz-invariant) ground state.
For the gravitinos this amounts to \cite{dWHS} 
 \begin{equation}\label{killingspinor1}
  \vev{\delta\psi_{\mu}^I}=\vev{\nabla_{\mu}\epsilon^I+igA_1^{IJ}
  \gamma_{\mu}\epsilon^J}=0\ .
 \end{equation} 
In order to solve this Killing spinor equation for an AdS or Minkowski
ground state, it is sufficient to make
a product ansatz for $\epsilon^I$ in terms of an eigenvector of $A_1$
and a three-dimensional AdS/Minkowski Killing spinor \cite{CGP}. 
Using  $ [\nabla_{\mu},\nabla_{\nu}]\epsilon 
=\frac{i}{2}R_{\mu\nu ab}\gamma^{ab}\epsilon$
and $R_{\mu\nu\rho\sigma} = 4V_0(g_{\mu\rho}g_{\nu\sigma}
-g_{\rho\nu}g_{\mu\sigma})$ we see, that 
eq.\ (\ref{killingspinor1}) can only be solved if
the following integrability relation is satisfied
 \begin{equation}\label{groundstate}
  g|\lambda|=\sqrt{-V_0}\ .
 \end{equation} 
Here $\lambda$ is an eigenvalue of  $A_1$ 
and $V_0$ denotes the cosmological constant, 
i.e.  the value of the potential 
in the ground state. 
Thus each eigenvector of $A_1$ 
whose eigenvalue satisfies (\ref{groundstate}) 
yields
a solution of the Killing spinor equation 
(\ref{killingspinor1}). 

The supersymmetry transformations of the other fermions 
$\chi^{\ir{i}}$ evaluated in the ground state are given by \cite{dWHS}
 \begin{equation}\label{killingspinor2}\,
  \vev{\delta\chi^{\ir{i}}}=-2g\vev{g^{\ir{i}\ir{j}}A_{2\ir{j}}^{I}}\, 
  \epsilon^I.
 \end{equation}
We see 
from (\ref{killingspinor1}) and (\ref{killingspinor2})
that each conserved supersymmetry corresponds to a spinor parameter 
$\epsilon^I$ which is a common eigenvector of $A_1$ and $A_{2}$, 
for $A_1$ with an eigenvalue related to the 
cosmological constant by 
(\ref{groundstate}) and for $A_2$ 
with a zero eigenvalue.\footnote{Strictly speaking, $A_2$ cannot have an 
eigenvector, because it is a rectangular matrix. 
However, we just mean by this 
the statement $A_{2\ir{j}}^I\epsilon^I=0$.}  
Using the  identity (\ref{quad2}) 
one infers that both conditions or in other words
(\ref{killingspinor1}) and (\ref{killingspinor2})
are necessarily satisfied simultaneously. 
Put differently, each eigenvalue of 
$A_1$ satisfying (\ref{groundstate}) 
is automatically an eigenvector of $A_2$ with
vanishing eigenvalue.
Therefore, one only has to determine the eigenvectors
of $A_1$ with eigenvalue (\ref{groundstate}) 
in order to find the unbroken supercharges.
Using (\ref{a1}) one obtains 
 \begin{equation}\label{eigen}
  \lambda_{\pm}=-2T\pm e^{K/2}|W|
  =-\frac{1}{2}M_A\Theta^{AB}M_B \pm e^{K/2}|W|\ ,
 \end{equation}
where the second equation used  (\ref{T}).
Inserted into (\ref{groundstate}) and using  
(\ref{pot2}) one arrives at
 \begin{equation}\label{rel2}
  \begin{split}
    \pm 2e^{K/2}M_A\Theta^{AB}M_B|W|
    =\ &M_A\Theta^{AC}G_{CD}\Theta^{DB}M_B+e^K|W|^2 M_A G^{AB}M_B \\
    &+\frac{1}{2}e^K G^{a\bar{b}}D_a W D_{\bar{b}}\overline{W}.
  \end{split}
 \end{equation} 
For the ground state to respect the full
$N=2$ supersymmetry, this relation 
has to be satisfied for both eigenvalues
$\lambda_{\pm}$, i.e. for both signs in (\ref{rel2}).
Since $\Theta^{AB}$ and $G_{AB}$ are 
non-vanishing and $G_{AB}$ is positive definite
each term in (\ref{rel2}) has to vanish separately.
This implies the two solutions
 \beq\label{bre1}
\Theta^{AB}M_B= W= D_a W=0\ , \qquad 
\textrm{or}\qquad
M_A=D_a W=0\ , \quad W\neq 0\ . 
 \eeq
Both solutions correspond to stationary points 
of $V$ in that they satisfy 
$\partial_A V=\partial_a V = 0$ as can be checked
from the expression (\ref{pot2}).
The first solution has vanishing cosmological
constant since it satisfies $T=V_0=0$.
The second solution has $T=0, V_0 = - g^2 e^K |W|^2$
and hence a negative cosmological constant.

For  partial $N=2\rightarrow N=1$ breaking
the situation is more involved. For this case 
the condition (\ref{rel2}) should only be satisfied
for one eigenvalue but not both. Thus the right
hand side of (\ref{rel2}) cannot vanish or
equivalently the left hand side must be non-vanishing.
This implies
\beq
|W|\neq 0 \, \qquad \textrm{and}\qquad \Theta^{AB}M_B \neq 0\ .
 \eeq

Without additional input (\ref{rel2}) cannot be 
further simplified. 
However, if we impose a vanishing cosmological
constant of the $N=1$ ground state, i.e.\ $V_0=0$,
eqs.\ (\ref{pot2}) and (\ref{rel2}) imply
\begin{equation}\label{bed2}
  \pm 2e^{K/2}|W|=M_A\Theta^{AB}M_B\ ,
 \end{equation}
for precisely one choice of the signs.
In order to simplify the analysis we follow \cite{BHS} and 
assume $M_AG^{AB}M_B=2$. In this case the potential (\ref{pot2}) 
is manifestly positive-definite
  \begin{equation}\label{pospot}
   \begin{split}
    g^{-2}V=\ &\frac{1}{2}(M_A\Theta^{AC}-2TM_AG^{AC})G_{CD}
    (\Theta^{DB}M_B-2TG^{DB}M_B) \\
    &+\frac{1}{4}e^{K}G^{a\bar{b}}D_a WD_{\bar{b}}\overline{W}\ ,
   \end{split}
  \end{equation}
and a sufficient condition for the minimum is given by 
  \begin{equation}\label{bed1}
   \Theta^{DB}M_B-2TG^{DB}M_B=0\qquad \text{ and }\quad D_a W=0\ .
  \end{equation} 
For the positive-definite potential (\ref{pospot}) the two conditions 
(\ref{bed1}) and (\ref{bed2}) are necessary and sufficient for a 
$N=1$ Minkowski ground state \cite{BHS}.

\section{The $N=1$ Supermultiplets}
\setcounter{equation}{0}
In this section we study the spectrum of a spontaneously broken 
$N=2$ supergravity
or in other words the rearrangement of the $N=2$ multiplets in terms of 
$N=1$ multiplets. We only discuss the case of Minkowskian
ground states leaving the study of the AdS case to a separate
publication.

As we observed in the previous section an $N=1$ Minkowskian ground state 
is characterized by one zero and one non-zero eigenvalue of $A_1$.
{}From eq.\ \eqref{yukawa} we see that $A_1$ is the mass
matrix of the two gravitini and thus one zero and one non-zero eigenvalue
corresponds to a  massless and a  massive gravitino.
Which of the eigenvalues given in \eqref{eigen}
is zero is a matter of convention and 
without loss of generality we  choose to discuss the situation 
$\lambda_+\neq0, \lambda_- = 0$. 
We will see in the following that the gravitino mass is related to
the non-vanishing eigenvalue 
by $\lambda_+ = 2g^{-1} m_{\psi}$.
Together with \eqref{bed2} this implies 
\beq\label{gravimass}
m_{\psi}= \frac12 g\lambda_+=ge^{K/2}|W|=-\frac{1}{2}gM_A\Theta^{AB}M_B
= -2gT\ ,
\qquad \lambda_- = 0\ .
\eeq

A massive gravitino requires a super-Higgs effect or in other words
the presence  of a Goldstone fermion $\eta$ which by
an appropriate redefinition of the fields can be 
eliminated from the action (i.e.\ `eaten' by the gravitino).
In section \ref{susyhiggs} we discuss the details of this 
super-Higgs effect and in section
\ref{higgstop} we show how the supersymmetric partner
of the gravitino receives its mass by a topological Higgs mechanism.

\subsection{The super-Higgs effect in $D=3$}\label{susyhiggs}
In analogy with the situation in $D=4, N=1$ \cite{wess} let us define 
$\eta^I=\vev{A_{2\ir{j}}^{I}}\chi^{\ir{j}},\ I=1,2$. 
Using \eqref{killingspinor2}
one immediately infers the transformation law of $\eta$
(evaluated in the ground state)
to be
\begin{equation}\label{etatrans}
  \vev{\delta\eta^I}=-2g\vev{A_{2\ir{j}}^{I}g^{\ir{j}\ir{k}}
  A_{2\ir{k}}^{K}}\epsilon^K.
 \end{equation}
For the conserved supersymmetry $\epsilon^K$
is an eigenvector of $A_2$ with vanishing eigenvalue
and hence $\vev{\delta\eta^I}=0$ holds. For the broken supersymmetry 
we have instead  
 \begin{equation}\label{goldstonevar}
   \vev{\delta\eta^I}=-2g\vev{A_{2\ir{j}}^{I}g^{\ir{j}\ir{k}}
   A_{2\ir{k}}^{K}}\epsilon^K
   =-2g\vev{A_1^{IJ}A_1^{JK}}\epsilon^K 
   =-8g^{-1}m_{\psi}^2\epsilon^I\ ,
 \end{equation}   
where we used (\ref{quad2}) and (\ref{gravimass}).
Eq.\ \eqref{goldstonevar} shows that 
$\eta^I$ transforms inhomogeneously (by a shift) exactly as required 
by a Goldstone fermion.

In the ground state the matrix $A_2$ considerably simplifies
as can be seen by inserting \eqref{bed1} and \eqref{T}
into the definition given in \eqref{a2}. 
Since  $T$ is only a function of the $M_A$ 
we see that in the direction of the chiral multiplets
$\vev{A_{2a}}=0$ holds while in the direction of
the vector multiplets we have 
$\vev{A_{2A}}\neq0$. This implies that the
Goldstone fermion $\eta^I$ is a linear combination of only the fermions
in the vector multiplets but does not contain any fermions
in chiral multiplets.

Exactly as in $D=4,N=1$ it is possible to redefine the massive 
gravitino and absorb the Goldstone fermion. 
Let us first perform an $SO(2)$ transformation on the $\psi_{\mu}^I$ 
such that $A_1^{IJ}$ in (\ref{yukawa}) is  diagonalized. 
Since $A_1$ and $A_2$ have 
common eigenvectors we can rotate $\eta^I$ accordingly.
In this rotated basis $\eta^1$ vanishes from the Lagrangian since
$A_{2\ir{i}}^1=0$ 
holds as a consequence of the conserved supersymmetry.\footnote{Note that
in the case that both supersymmetries are broken also $A_{2\ir{i}}^1$ 
would be different from zero, corresponding to the
second Goldstone fermion.}  
Thus we arrive at 
 \bea\label{yukawa2}
  e^{-1}\mathcal{L}_Y =
  m_{\psi}{\bar{\psi}}_{\mu}^2\gamma^{\mu\nu}{\psi}_{\nu}^2
  +2g{\bar{\psi}}_{\mu}^2\gamma^{\mu}{\eta^2}
  +2igA_{3\ir{i}\ir{j}}\bar{\chi}^{\ir{i}}\chi^{\ir{j}}. 
 \eea
As expected the massless gravitino $\psi_{\mu}^1$ disappeared 
from the mass terms
and we are left with an off-diagonal fermionic mass matrix.
Before diagonalizing the Yukawa coupling let us split off the 
physical spin-$1/2$ fermions according to
 \begin{equation}
  \chi_{\bot}^{\ir{i}}
  :=\chi^{\ir{i}}-\frac{g^2}{\sqrt{2}m_{\psi}^2}
   \vev{g^{\ir{i}\ir{j}}A_{2\ir{j}}^2}\eta^2,
 \end{equation} 
such that the fermionic part of the action looks together with 
the kinetic terms like
\bea\label{yukawa3}
  e^{-1}\mathcal{L}_F &=&
  -\frac{1}{2}i\varepsilon^{\mu\nu\rho}\bar{\psi}^2_{\mu}
  \partial_{\nu}\psi^2_{\rho} 
  -\frac{1}{2}g_{\ir{i}\ir{j}}\bar{\chi}_{\bot}^{\ir{i}}
  \gamma^{\mu}\partial_{\mu}\chi^{\ir{j}}_{\bot}
  -\frac{g^2}{m_{\psi}^2}\bar{\eta^2}\gamma^{\mu}\partial_{\mu}\eta^2 \\
  &&+
  m_{\psi}{\bar{\psi}}_{\mu}^2\gamma^{\mu\nu}{\psi}_{\nu}^2
  +2g{\bar{\psi}}_{\mu}^2\gamma^{\mu}{\eta^2}
  +2igA_{3\ir{i}\ir{j}}\bar{\chi}^{\ir{i}}\chi^{\ir{j}}, 
  \nonumber 
 \eea
where we have used (\ref{quad2}). 
By an appropriate redefinition of the massive gravitino
$\psi^2_{\rho}$ the Goldstone fermion $\eta^2$ can be removed from
the entire action. This redefinition  
is inspired by the supersymmetry transformations 
of $\eta$ given in \eqref{etatrans}
with $\epsilon^I=\frac{g}{8m_{\psi}^2}\eta^I$, including 
a term proportional to $\partial_\mu \eta$ in order to 
remove the kinetic term for the Goldstone fermion. 
It reads (omitting the index 2)  
\begin{equation}\label{psiredef}
  \hat{\psi}_{\mu}=\psi_{\mu}+\frac{g}{m_{\psi}^2}\partial_{\mu}\eta
  -\frac{g}{m_{\psi}}i\gamma_{\mu}\eta\ ,
 \end{equation}
resulting in 
 \begin{equation}\label{supermasse}
  e^{-1}\mathcal{L}_F =
  -\frac{1}{2}i\varepsilon^{\mu\nu\rho}\hat{\bar{\psi}}_{\mu}
  \partial_{\nu}\hat{\psi}_{\rho}
  -\frac{1}{2}g_{\ir{i}\ir{j}}\bar{\chi}_{\bot}^{\ir{i}}
  \gamma^{\mu}\partial_{\mu}\chi^{\ir{j}}_{\bot}+
  m_{\psi}\hat{\bar{\psi}}_{\mu}\gamma^{\mu\nu}\hat{\psi}_{\nu} 
  + i(m_\chi)_{\ir{i}\ir{j}} \bar{\chi}^{\ir{i}}\chi^{\ir{j}}\ ,
 \end{equation}   
where 
\beq\label{fermass}
(m_\chi)_{\ir{i}\ir{j}} =2gA_{3\ir{i}\ir{j}} 
-\frac{3g^2}{m_{\psi}}A_{2\ir{i}}A_{2\ir{j}}\ .
\eeq
As promised $\eta$ has disappeared from the action 
and left a zero eigenvalue in the mass matrix 
$(m_\chi)_{\ir{i}\ir{j}}$. This zero eigenvalue is most easily seen
by observing that $g^{\ir{j}\ir{k}}A_{2\ir{k}}$ is a null vector
of (\ref{fermass}) 
 \bea
  m_{\ir{i}\ir{j}}\, g^{\ir{j}\ir{k}}A_{2\ir{k}}&=&
  2g(A_{3\ir{i}\ir{j}}g^{\ir{j}\ir{k}}A_{2\ir{k}}
  +\frac{3g}{2m_{\psi}}A_{2\ir{i}}g^{\ir{j}\ir{k}}A_{2\ir{j}}A_{2\ir{k}}) \\
  &=& 2g(g^{\ir{j}\ir{k}}A_{3\ir{i}\ir{j}}A_{2\ir{k}}
  +\frac{6m_{\psi}}{g}A_{2\ir{i}})=0, \nonumber
 \eea 
where we used (\ref{a1}) and (\ref{quad2}) together with (\ref{gravimass})
and (\ref{quad3}). The use of (\ref{quad3}) is justified by
the results of \cite{CGP} in that a $N=1$ configuration
is necessarily a stationary point of the potential. 
We see that after the redefinition \eqref{psiredef}
the Goldstone fermion disappeared from the action 
and left a properly
normalized massive gravitino with one 
physical degree of freedom. This propagating degree of freedom
can be seen 
by applying the differential operator 
$(-i\varepsilon_{\mu\lambda\sigma}\partial^{\lambda}-2m_{\psi}
\gamma_{\mu\sigma})$ to 
the equation of motion for the redefined gravitino
(we are dropping the `hats' henceforth)
 \begin{equation}\label{eom}
 i\varepsilon^{\mu\nu\rho}\partial_{\nu}\psi_{\rho}
  -2m_{\psi}\gamma^{\mu\nu}\psi_{\nu} = 0 \ .
 \end{equation}   
Up to linear terms in the derivatives 
this results in 
 \begin{equation}
   0=\left(\square +m_{\psi}^2\right)\psi_{\mu} -\partial_{\mu}
   (\partial\cdot\psi) + \ldots \ ,
 \end{equation}
which shows  that the gravitino has become a massive propagating 
field. Altogether it
carries now one fermionic degree of freedom.\footnote{Counting the 
physical degrees of freedom 
is also consistent with the reduction from $D=4$.
In $D=4$ a massive gravitino has four degrees of 
freedom and it splits in the reduction as $\psi_{m\alpha}=
(\psi_{\mu\alpha},\psi_{3\alpha})=(\psi_{\mu\alpha}^1+i\psi_{\mu\alpha}^2,
\psi_{3\alpha}^1+i\psi_{3\alpha}^2)$, where $m=0,\ldots,3$.
{}From a three-dimensional point of view the real and imaginary parts 
of $\psi_m$ are real Majorana 
spinors of the $D=3$  Lorentz group 
$SL(2,\mathbb{R})$. Thus we see that each of the 
$(\psi_{\mu\alpha}^1, \psi_{\mu\alpha}^2,
\psi_{3\alpha}^1,\psi_{3\alpha}^2)$ carries one  degree of freedom.}
Thus the originally massless and topological gravitino
(with no physical degree of freedom) becomes 
propagating due to the presence of the mass term.
The physical degree of freedom of the massive gravitino
coincides with the physical degree of freedom of
the Goldstone fermion it has eaten.
A similar situation occurs in the
topological Higgs mechanism for vector fields
which we turn to now. 

\subsection{The topological Higgs mechanism}\label{higgstop}
The spontaneous supersymmetry breaking we are considering in this paper
leaves an $N=1$ unbroken. Therefore, after the breaking the original $N=2$
multiplets have to rearrange in $N=1$ multiplets. This 
means in particular that the massive gravitino must be the member
of a massive $N=1$ multiplet and in this section we
identify the supersymmetric partner of the massive gravitino.
In appendix B we review the general structure of the 
massive $N=1$ supermultiplets in $D=3$ and show that the irreducible 
multiplets contain the spins $(j,j+\frac{1}{2})$. 
Because a massive gravitino even in $D=3$ has spin $\frac{3}{2}$
\cite{superdeser}, we expect
a massive vector to be the supersymmetric partner of the massive gravitino.
This is also suggested by the fact that the Goldstone fermion
arises solely from the vector multiplets leaving the chiral multiplets
untouched. Therefore the supersymmetric partner of the massive gravitino
should also come out of the vector multiplets. 

However, this proposition is somewhat puzzling at first sight.
The reason is that as we just argued the massive gravitino has only
one physical degree of freedom which is the same as a massless vector.
In the standard Higgs mechanism the vector acquires a mass 
by `eating'  a (spin-0) Goldstone boson
which adds one degree of freedom to the vector.
A second problem is that in the action (\ref{wirkung}) there is no 
candidate for a Goldstone boson which has the appropriate couplings.
The resolution of this apparent paradox is the possibility of a 
topological Higgs effect which does exist in $D=3$ \cite{schonfeld,deser1}.
Let us briefly review the mechanism.

This topological Higgs effect arises when one adds a Chern-Simons
term to the standard Yang-Mills term or in other words the 
gauge invariant action is given by
 \begin{equation}\label{topo}
  \mathcal{L}=\frac{1}{4}F_{\mu\nu}F^{\mu\nu} - \frac{1}{2}\xi
  \varepsilon^{\mu\nu\lambda}A_{\mu}\partial_{\nu}A_{\lambda}\ .
 \end{equation}
From the equations of motion one shows that the theory defined by
\eqref{topo} has a massive excitation with mass $\xi$ 
\cite{schonfeld,deser1}. The proof is analogous
to the argument presented at the end of the previous section
for the gravitino. The equation of motion derived from the action 
\eqref{topo} reads 
 \begin{equation}\label{eofA}
 \xi F^{\mu}-\varepsilon^{\nu\lambda\mu}\partial_{\nu}F_{\lambda} = 0\ ,
 \end{equation}
where $F^{\mu} \equiv\varepsilon^{\mu\nu\rho} F_{\nu\rho}$.
Applying to \eqref{eofA} the first-order 
differential operator $\xi g_{\sigma\mu}+\varepsilon_{\sigma\rho\mu}
\partial^{\rho}$ and using the Bianchi identity $\partial_{\mu}F^{\mu}=0$
one derives\footnote{Note that the 
sign in front of the Chern-Simons term 
is in view of the massive excitation arbitrary, 
because it cannot affect the quadratic mass term in \eqref{KG}.  
But, according to appendix B, the spin could be +1 or -1 and 
which one is realized is determined by this sign \cite{dWNS}.} 
\beq\label{KG}
 (\square + \xi^{2})F_{\mu}=0,
\eeq
which proves the massive excitation. 
Since no Goldstone boson was eaten,  $A_\mu$ 
still has only one physical
degree of freedom and furthermore the gauge invariance is unbroken.

The action \eqref{wirkung} includes the two terms given in 
\eqref{topo}. Thus it possibly generates at least one massive vector,
because the gauge coupling functions $G^{AB}$ get a 
non-vanishing vev. 
The only thing left to show is that among the massive vectors
there is one which is degenerate with the gravitino and has a mass $m_\psi$.
In order to display this vector we need to canonically normalize
the Yang-Mills action in the ground state.
This can be achieved by rotating the vector fields
according to an $SO(n)$-transformation 
$A_{\mu A} \to S_{A}^{\hspace{0.5em}B}A_{\mu B}$ such that the $G^{AB}$ are 
diagonalized and normalized to be the identity
matrix. In this basis 
the physical mass matrix for the vector fields is given in view of 
(\ref{topo}) by
\beq\label{spin1mm}
m^{AB} =  -g\, (S^{-1}\Theta S)^{AB} \ .
\eeq
Now we need to show that $m^{AB}$ has at least one eigenvalue
given by $m_\psi$. It can be most easily seen by multiplying
\eqref{spin1mm}
 by $S^{-1}_{BC} M_C$ and using \eqref{bed1}. This implies
\bea\label{mcomp}
m^{AB}S^{-1}_{BC} M_C 
&=&  -g\, (S^{-1}\Theta)^{AC} M_C = -2g T (S^{-1}G)^{AC} M_C \nonumber\\
&=& -2g T (S^{-1}GS)^{AB} S^{-1}_{BC}M_C = -2g T  S^{-1}_{AB}M_B\ . 
\eea
This equation indeed shows that $S^{-1}_{BC} M_C $ is an eigenvector
of $m^{AB}$ with eigenvalue $-2gT=m_\psi$ (here we used 
\eqref{gravimass}). Thus the mass matrix \eqref{spin1mm} for the vectors
has at least one eigenvalue  $m_\psi$ as required by the unbroken $N=1$ 
supersymmetry.

Before we continue let us note that this discussion can also be carried out
in terms of the original Lagrangian \eqref{boson}
where no Yang-Mills kinetic term is present but instead a set
of charged scalar fields $\hat \varphi^A$ with covariant
derivatives given in \eqref{covariant2}. 
In this field basis the Chern-Simons term acts as a kinetic term
and the  $\hat \varphi^A$ are the Goldstone bosons giving a mass
for the gauge fields. (Compare with the discussion in ref.\ \cite{FNS}.)
This version of the topological Higgs 
mechanism has also been discussed in the literature \cite{jackiw,townsend}
and we will briefly review it.  

Starting from the action
 \begin{equation}\label{topo2}
  \mathcal{L}=\frac{1}{2}\varepsilon^{\mu\nu\rho}A_{\mu}
  \partial_{\nu}A_{\rho}-\frac{1}{2}\xi A_{\mu}A^{\mu}
 \end{equation}
with a conventional mass term, also leads
to a massive excitation of one propagating degree of freedom 
\cite{townsend}.
Namely the equation of motion states that the vector field is 
Hodge-dual to the field strength, 
$A_{\mu}=-\frac{1}{2\xi}\varepsilon_{\mu\nu\rho}F^{\nu\rho}$,
which in turn implies the Proca equation for a massive vector
 \begin{equation}
  \partial^{\mu}F_{\mu\nu}+\xi^2A_{\nu}=0,
 \end{equation}
together with the Lorentz condition $\partial_{\mu}A^{\mu}=0$. 
Furthermore, (\ref{topo}) and (\ref{topo2}) are equivalent in 
the sense that both can be derived from a ``master Lagrangian'' by varying 
different fields \cite{jackiw}. But in this version it is possible 
to relate this topological effect in a more standard way to the
notion of spontaneous symmetry breaking. 

In the original action (\ref{boson}), where no scalar fields have 
been dualized, the relevant part can be written as
 \bea\label{topo3}
  \mathcal{L}&=&\frac{1}{2}\,G(M)\,\mathcal{D}_{\mu}\hat\varphi 
  \mathcal{D}^{\mu}\hat\varphi
  +\frac{1}{2}\,g\, \Theta\, \varepsilon^{\mu\nu\rho}A_{\mu}\partial_{\nu}A_{\rho}
\nonumber\\
  & =&\frac{1}{2}\, g^2\Theta^2\, G(M)\, A'_{\mu}A^{\prime\mu}
   +\frac{1}{2}g\,\Theta\,\varepsilon^{\mu\nu\rho}A'_{\mu}\partial_{\nu}
   A'_{\rho}\ ,
 \eea
where $\mathcal{D}_{\mu}\hat\varphi = \partial_\mu\hat\varphi + g\Theta A_\mu$
and for simplicity we have restricted the discussion to 
one vector field.
In the second equation the scalar field has been absorbed by
a gauge transformation into the gauge field, 
leaving only a (St\"uckelberg) mass term for $A_\mu$.
Due to (\ref{gravimass}) we have
$-g\Theta\vev{G(M)}=m_{\psi}$, such that $A_{\mu}$ gets the 
gravitino mass. 
Hence $\hat\varphi$ plays the role of the standard
Goldstone boson which provides one physical degree of freedom 
to a previously topological gauge field.
Note that the potential does not depend on $\hat\varphi$, which means
that $\hat\varphi$  remains massless with respect to an arbitrary 
groundstate, as required for Goldstone bosons.  
Also in this field basis one shows in general that the mass matrix 
has at least one eigenvalue given by $m_\psi$. 

Let us note that this situation is a little bit different from 
the one considered in \cite{deser2}.
In our case the PQ-isometries form a non-compact gauge group
isomorphic to $\mathbb{R}$ acting only
on the real scalar fields $\hat\varphi$.
In contrast one can start like in \cite{deser2}
from a Chern-Simons $U(1)$-gauge theory coupled to a
complex scalar and break the gauge invariance by adding a usual 
Higgs potential. Gauging the phase factor of the complex scalar away, 
the action reduces to (\ref{topo2}), together with the kinetic term 
for the left-over real Higgs field. Therefore the phase 
factor plays the role of the Goldstone boson, which is absorbed 
into the gauge field by a phase transformation. 
In this setup one has the standard connection of the mass of a gauge boson 
with spontaneous symmetry breaking.

\subsection{The $N=1$ mass spectrum}

We started our analysis from a generic $N=2$ spectrum 
given in \eqref{spectrum} containing a gravitational multiplet,
$n$ vector multiplets
and $d-n$ scalar multiplets. Let us now show 
how after spontaneous $N=2\to N=1$
breaking this spectrum arranges into $N=1$ multiplets.
In $D=3$ the possible irreducible massive or massless 
$N=1$ multiplets are 
a chiral multiplet $(\varphi,\chi)$
containing one real scalar $\varphi$ and one real Majorana fermion $\chi$,
a vector multiplet  $(A_\mu,\chi)$ containing a vector
$A_\mu$ and a Majorana fermion, the massive gravitino multiplet 
$(\psi_\mu, A_\mu)$
containing a massive gravitino $\psi_\mu$
and a massive vector $A_\mu$ and the gravitational multiplet
$(g_{\mu\nu},\psi_\mu)$ containing the metric and a massless gravitino.
All multiplets have one fermionic and one bosonic degree of freedom
except that the gravitational multiplet contains no degrees of freedom.

From this representation theory and the consideration of
the previous section we see that
after partial supersymmetry breaking the original $N=2$
spectrum assembles into one gravitational multiplet,
one massive gravitino multiplet, $n-1$ vector multiplets,
$n$ chiral multiplets coming out of the $N=2$ vector multiplets
and $2(d-n)$ chiral multiplets from the sector of complex scalars. 
More precisely we have 
\bea\label{spectrum1}
&&[g_{\mu\nu},\psi^1_{\mu},\psi^2_{\mu}]\
  \oplus\ n\times [A_{\mu},\chi^1,\chi^2,M_A]\ \oplus\
 (d-n)\times [\phi,\chi_c]\ \longrightarrow\\
&&\qquad
[g_{\mu\nu},\psi^1_{\mu}]\ \oplus \
  [\psi^2_{\mu},A_{\mu}]\
  \oplus\ (n-1)\times [A_\mu,\chi]\ \oplus\ n\times [M,\chi] \
\oplus\ 2(d-n) \times [\varphi,\chi] \ .\nonumber
 \eea

We already argued that the gravitino multiplet is massive
and so we are left to determine the masses of the other
vector- and chiral multiplets.

For the vectors we already recorded their mass matrix in 
\eqref{spin1mm} and the number of massive vector fields will be
given by the rank of $\Theta^{AB}$.  
For the scalar fields coming out of the vector multiplets 
one computes the mass matrix
as the second derivative of the scalar potential, 
$m^{AB} = \partial_{M_A}\partial_{M_B} V$.   
We focus on the 
positive-definite potential (\ref{pospot}) and 
first note that in this case
the condition $M_AG^{AB}M_B=2$ implies 
that $\partial_a\partial_{M_B}K=0$. This again shows together
with $W=W(\phi^a)$ that the K\"ahler-covariant 
derivative $D_aW$ is independent of $M_A$. 
As a consequence  the mixed derivative of (\ref{pospot}) with respect to 
$M_A$ and $\phi^a$ vanishes in view of (\ref{bed1}),
i.e.\ the mass spectrum decouples between the scalars belonging 
to the chiral and the vector multiplets.
We will not reproduce the full expression for $m^{AB}$, but 
just state that depending on the geometry given by $G_{AB}$ 
some of the $M_A$ remain massless
and some of them become massive with masses which are of the
order of $m_{\psi}$.  
 
For the fermions belonging to the scalar multiplets
$[M_A,\chi_A]$ as well as the fermions of the vector multiplets
$[A_{\mu},\chi]$ the masses
can be computed from $A_3$ and $A_2$ using (\ref{fermass}) 
upon making the coordinate transformation to $(\phi^a,M_A)$.
However, we will not work out the details here. 

Let us now turn to the chiral multiplets $[\varphi,\chi]$.
As we have seen there is no coupling between the scalars belonging to 
vector and scalar multiplets, such that the 
mass matrix of the $\phi^a$ is determined by the second derivatives
 \begin{equation}\label{phimass}
\begin{split}
   M_{ab}^2 &=
   \vev{\frac{\partial^2V}{\partial\phi^a\partial\phi^b}}
   =\frac{g^2}{4}\, e^K\overline{W}
   \left[D_a(D_bW)+D_b(D_aW)\right]\ ,\\
  M_{a\bar{b}}^2 &=
  \vev{\frac{\partial^2V}{\partial\phi^a\partial\bar{\phi}^{\bar{b}}}}
  =\frac{g^2}{4}\, e^K\left[g^{c\bar{d}}D_a(D_cW)D_{\bar{b}}
  (D_{\bar{d}}\overline{W}) + g_{a\bar{b}}|W|^2 \right]\ ,
 \end{split}
\end{equation}
where we used (\ref{bed1}).


As a next step we calculate the corresponding fermion masses. 
We first note that as for the scalar fields there is no coupling 
between the fermions belonging to the vector multiplets and the ones
belonging to the scalar multiplets. 
This can be seen by inspecting $A_{3aA}$ in (\ref{a2}) which shows that
it is proportional to $D_aD_AW\sim M_A D_{a}W$
and which vanishes in the ground state due to (\ref{bed1}).
Altogether, we see that the rearrangement of the multiplets 
runs independently between the vector and scalar 
multiplets, as indicated by (\ref{spectrum1}).\footnote{Note that 
this is only true for the 
positive-definite potential 
(\ref{pospot}) and the corresponding conditions (\ref{bed1}).
In general the scalar and vector multiplets can couple
in a complicated way in which case the analysis would be much more involved.} 
  
The masses of the fermionic partners of the $\phi^a$  are 
directly given by $A_{3\ir{a}\ir{b}}$ since  these fermions 
do not mix with the Goldstone fermion (see \ref{susyhiggs}) and thus
there is no coupling with the gravitinos. 
From (\ref{fermass}) we learn that the unnormalized
mass matrix of these fermions is given by 
$m_{\ir{a}\ir{b}}=2g\vev{A_{3\ir{a}\ir{b}}}$. With (\ref{a2}) these matrices
can be written in a complex notation as 
 \begin{equation}
   m_{ab}=\frac{g}{2}e^{K/2}D_a(D_bW)\ , \qquad
   m_{a\bar{b}}=-gg_{a\bar{b}}T=\frac{g}{2}e^{K/2}g_{a\bar{b}}|W|\ ,
 \end{equation}
where we have used for the second equation (\ref{bed2}).
To compare these mass matrices with the ones for the scalars, we compute
their squares $\mathcal{M}_{\ir{i}\ir{j}}^2=g^{\ir{k}\ir{l}}m_{\ir{i}\ir{k}}
m_{\ir{j}\ir{l}}$ and
arrive at
 \begin{equation}\label{chimass}
  \begin{split}
   \mathcal{M}_{ab}^2&=g^{d\bar{c}}m_{ad}m_{b\bar{c}}+
   g^{c\bar{d}}m_{a\bar{d}}m_{bc} 
   =\frac{g^2}{4}e^K\overline{W}\left[D_a(D_bW)+D_b(D_aW)\right]\ ,\\
   \mathcal{M}^2_{a\bar{b}}&=g^{c\bar{d}}m_{ac}m_{\bar{b}\bar{d}}
   +g^{d\bar{c}}m_{a\bar{c}}m_{\bar{b}d} 
   =\frac{g^2}{4}e^K\left[g^{c\bar{d}}D_a(D_cW)D_{\bar{b}} 
   (D_{\bar{d}}\overline{W}) + g_{a\bar{b}}|W|^2\right]\ ,
  \end{split}
 \end{equation}
where in the first equation we have 
performed an appropriate K\"ahler transformation to write $\overline{W}$
instead of $|W|$.
As demanded by 
$N=1$ supersymmetry we find mass degenerate chiral multiplets or 
in other words agreement between (\ref{phimass}) and
(\ref{chimass}).

\subsection{The $N=1$ effective action}
Well below the scale of partial $N=2\to N=1$
supersymmetry breaking set by $m_{\psi}$ the dynamics 
of the light $N=1$ multiplets can be best described by an
effective action $\mathcal{L}_{\text{eff}}^{N=1}$. 
Since $N=1$ is unbroken this 
action should be manifestly $N=1$ supersymmetric.
It is calculated by 
``integrating out'' the massive gravitino multiplet together
with all other multiplets with masses of order $m_{\psi}$.  
The resulting effective action can be obtained as a
power series expansion in  $p/m_{\psi}$ where $p$ is a
typical momentum scale satisfying $p\ll m_{\psi}$.
To lowest non-trivial order in $p/m_{\psi}$ this amounts to
setting all massive multiplets equal to zero
keeping only the left-over light $N=1$ multiplets.
In effect this truncates the scalar manifold to a subspace
of the original K\"ahler manifold and projects out a set
of heavy vector multiplets.
Due to the topological nature of the Higgs mechanism in $D=3$
this is very different from the situation
in $D=4$ where the mixing of the Goldstone bosons 
results in taking a certain quotient space of the original
scalar manifold \cite{louis1}. Here the scalar manifold
is merely truncated to a submanifold and one is left
with a standard (ungauged) $N=1$ supergravity action
coupled to light chiral multiplets.

Let us further note that due to this truncation 
the geometry loses generically some of its structure.
According to the fact that the $N=1$ multiplets contain 
at most one real scalar, setting the massive multiplets 
equal to zero in general distinguishes between real 
and imaginary parts. As a consequence the holomorphicity 
property of the superpotential as well as the complex 
structure of the scalar manifold is lost. 
But, this is in turn consistent with $N=1$ supersymmetry 
which only demands the scalar manifold to have a Riemannian 
structure and the superpotential to be a real 
function \cite{dWHS}.

\section{Conclusions}
\setcounter{equation}{0}

In this paper we analyzed models of partial $N=2\to N=1$ breaking
in three-dimensional $N=2$ supergravity. These models are inspired by
string compactifications of M-theory on Calabi-Yau fourfolds \cite{BHS}
and correspond to supergravities where translational (Peccei-Quinn)
isometries are gauged.
We saw that the massive gravitino multiplet consists of
a gravitino and a vector boson degenerate in mass which both carry
one degree of freedom. This is possible in $D=3$ due to the 
topological Higgs mechanism where the gauge boson mass is
generated by a Chern-Simons term without the necessity of any Goldstone boson.
Alternatively, in a dual description this topological Higgs mechanism
can be described with a standard coupling to a
Goldstone boson, which was precisely the scalar field that could
be dualized into the vector before a spontaneous breaking.  
But in this field basis, where all degrees of freedom reside in the
scalar sector, the Yang-Mills kinetic term is absent and instead 
only a topological Chern-Simons terms is present. 

We argue now that in $D=3$ the presented picture is in general valid.
Indeed, it was shown in \cite{NS} that gauged supergravity with 
Chern-Simons kinetic term on the one hand and 
Yang-Mills kinetic term on the other hand (Yang-Mills or Chern-Simons
gauging) are equivalent in the following sense.
Using the on-shell duality between scalars and vectors, a Chern-Simons 
theory with gauge group $G^{\prime}=G\ltimes T$, 
where $T$ denotes
an abelian group of translations transforming non-trivially under $G$,
can be transformed into a
Yang-Mills theory with gauge group $G$, i.e. in the Yang-Mills
picture the gauge group is broken to a smaller one.\footnote{In the
model discussed here we had the situation that $G$ is trivial, i.e.
that there are no charged scalars in the Yang-Mills picture.}
Therefore they are now two ways of obtaining massive vector fields
with respect to a non-trivial groundstate.
In the picture of pure Chern-Simons gauging the subgroup
$T$ of the gauge symmetry is broken, leading to massive vectors via 
eating the Goldstone bosons corresponding to the broken symmetry
in analogy with (\ref{topo2}). Otherwise, in the picture of
Yang-Mills gauging the gauge group is still the subgroup 
$G\subset G^{\prime}$, that means the gauge symmetry remains unbroken 
giving rise to topologically massive vectors corresponding 
to (\ref{topo}). In particular no Goldstone bosons get eaten 
because the corresponding degrees of freedom are still dualized 
into the gauge fields.\footnote{Compare the situation in the context of 
Kaluza-Klein supergravity analyzed in \cite{SN}.}

The class of models analyzed in this paper
contained gauged translational
isometries and only Chern-Simons kinetic terms for the gauge fields
and thus the possibility of a topological Higgs mechanism
existed right from the beginning. However, our analysis
of the super-Higgs effect suggests that this is of more general validity.
Since the massive $N=1$ gravitino multiplet
has one fermionic and one bosonic degree of freedom,
an ordinary Higgs mechnism which raises the number
of degrees of freedom of a gauge boson from one to two
is not suitable.

Following this line of reasoning we would like to clarify
the following characteristic property of supersymmetric field 
theories in $2+1$ dimensions, in order to consider further 
constraints on different scenarios of supersymmetry breaking. 
As it is explained in appendix B, the massive spin in $D=3$ is 
roughly the same as helicity in $D=4$ and the massive 
supermultiplets in $D=3$ are formally the same as the massless 
ones in $D=4$. In the latter case massless fields of arbitrary spin $s$ 
carry always two degrees of freedom, corresponding  to the 
two helicity states $\pm s$. Depending on the spin and the amount 
of supersymmetry one has therefore possibly to double
the massless supermultiplets to organize the fields of a 
supersymmetric theory into 
(no longer irreducible) representations of the 
superalgebra \cite{wess}, e.g. for an $N=1$ chiral multiplet
one has to take $(0,\frac{1}{2})\oplus (-\frac{1}{2},0)$. 
This is in contrast 
to the massive case in $D=3$. 
Here, e.g., a scalar multiplet 
consists of a real scalar and a real Majorana fermion, without the
need to double any multiplets.  
For the fermions this reduction of degrees of freedom 
is not surprising, because a complex Weyl spinor in $D=4$ 
splits in a Lorentz-invariant fashion into its real parts 
(compare sec. \ref{susyhiggs}).
On the other hand, if one simply takes the dimensional reduction
of massive bosonic theories in $D=4$, 
the fields in $D=3$ would also 
have more than one degrees of freedom and therefore would not be
well adapted for supersymmetry. 
(Compare the standard massive spin 1 action, which leads to two
degrees of freedom \cite{townsend}.)
Happily, as we have seen, 
with the topological mass term there exists a kind of 
``square root'' action for a massive vector,
which turned out to be necessary for supersymmetry.
Furthermore, in $D=3$ there is also a topologically massive spin-$2$ 
extension of gravity \cite{deser1} and a spin-$\frac{3}{2}$ extension of 
the Rarita-Schwinger field \cite{superdeser}, which together 
constitute topologically massive supergravity \cite{deser3}.   
In addition, this possibility of describing massive fields 
in an irreducible way, giving them one degree of freedom,
has been generalized to fields of arbitrary high spin
\cite{vasiliev}. 
One sees that these type of actions are much more natural 
for supersymmetry in $D=3$ and one should expect them to be 
of general importance. 
Keeping this in mind, similar partial breakings of 
supersymmetry or supergravity could be analyzed 
using the massive supermultiplets constructed in 
appendix B.

\subsection*{Acknowledgments}

This work is supported by DFG -- The German Science Foundation,
the European RTN Programs HPRN-CT-2000-00148, HPRN-CT-2000-00122,
 HPRN-CT-2000-00131 and  the
DAAD -- the German Academic Exchange Service.

We have greatly benefited from conversations with H.\ Jockers and especially
H.\ Samtleben.

J.L.\ thanks E. Cremmer and the L.P.T.E.N.S. in Paris for hospitality
and financial support during the final stages of this work.

\vskip 2cm

\noindent {\bf \LARGE Appendix}

\begin{appendix}
\renewcommand{\theequation}{\Alph{section}.\arabic{equation}}
  
\section{Spinor conventions}
\setcounter{equation}{0}
We use a space-time metric with signature 
$\eta_{\mu\nu}=\text{diag}(1,-1,-1)$ and a Clifford algebra representation 
given by
 \begin{gather}\label{darst}
  \gamma^0=-\sigma_2=\left(\begin{array}{cc} 0 & i \\ -i & 0 
  \end{array}\right)
  \text{,  }
  \gamma^1=i\sigma_3=\left(\begin{array}{cc} i & 0 \\ 0 & -i \end{array}\right)
  \text{,  }
  \gamma^2=-i\sigma_1=\left(\begin{array}{cc} 0 & -i \\ -i & 0 
  \end{array}\right).
 \end{gather}
Furthermore, we choose the charge-conjugation matrix to be
 \begin{equation}
  C=\sigma_2=\left(\begin{array}{cc} 0 & -i \\ i & 0 \end{array}\right),
 \end{equation}
such that the defining relation
 \begin{equation}\label{charge}
  C(\gamma^{\mu})^tC^{-1}=-\gamma^{\mu}
 \end{equation}
is satisfied and the Majorana spinors are the real
Dirac spinors. The Lorentz generators in the spinor representation are 
given by $\gamma^{\mu\nu}=\frac{i}{4}[\gamma^{\mu},\gamma^{\nu}]$
and the group generated by these is simply $SL(2,\mathbb{R})$, 
the double covering $\text{Spin}(1,2)$ of the Lorentz group. 

The important relations of the super algebra are given by
 \begin{align}
   \{Q_{\alpha}^I,\overline{Q}_{\beta}^J\}&=2(\gamma^{\mu})_{\alpha\beta}
   P_{\mu}\delta^{IJ}\label{susy}, \\
   [M^{\mu\nu},Q_{\alpha}^I]&=(\gamma^{\mu\nu})_{\alpha}^{\hspace{0.3em}\beta}
   Q_{\beta}^I\label{spinor},
 \end{align}
where the supercharges $Q_{\alpha}^I$ are Majorana spinors, 
$P_{\mu}$ denotes the 
energy-momentum operator and $M^{\mu\nu}$ the Lorentz generators. 
Using (\ref{charge}),
(\ref{susy}) can be written as
 \begin{equation}\label{susy2}
  \{Q_{\alpha}^I,Q_{\beta}^J\}=-2(\gamma^{\mu} C)_{\alpha\beta} 
  P_{\mu}\delta^{IJ}.
 \end{equation}

\section{Supermultiplets in $D=3$}
\setcounter{equation}{0}
In this appendix we review the massive and massless supermultiplets 
in $D=3$ for arbitrary $N$.
Let us first mention that it is not straightforward to use the spin 
as organizing principle in $D=3$. This is due to the fact that there is 
at least in the massless case no good concept of spin or helicity 
\cite{binegar}, which is reflected by the fact that half of the 
supercharges vanish. (See \cite{dWNT}, where the supermultiplets
in the massless case were constructed.) 
For completeness we give the massless multiplets in table 1,
where $d_n$ denotes the number of bosonic and fermionic degrees 
of freedom separately.
 
\begin{table}[h]\label{mult3}
  \begin{center}
   $\begin{array}{|c|c|c|c|c|c|c|c|c|c|   }
\hline
N   &  1 &  2 & 3 & 4 & 5 &  6 &   7 & 8 &n+8\\
\hline
d_n &  1 &  2 & 4 & 4 &  8 & 8 & 8 & 8 &16d_n\\
   \hline
   \end{array}$
   \caption{Massless Supermultiplets according to 
    \cite{dWNT}}
  \end{center}
 \end{table}

In order to write a standard Lagrangian one also needs supermultiplets 
with no physical degree of freedoms.
In the main text we use for $N=2$
$$
\begin{array}{llll}
\underline{N=2}\qquad& {\rm gravitational\ multiplet}\ & (g_{\mu\nu},\psi_{\mu}^I)\ &(0+0)\\
&{\rm vector\ multiplet}\        & (A_\mu, \lambda^I, M)\ & (2+2)\\
&{\rm chiral\ multiplet}\        & (\phi,\chi^I)\        & (2+2)\\
\end{array}
$$
where $I=1,2$. $\phi$ is a complex scalar while $M$ is a real scalar.
The fermions are all real Majorana.

For $N=1$ we use
$$
\begin{array}{llll}
\underline{N=1}\qquad& {\rm gravitational\ multiplet}\ & 
         (g_{\mu\nu},\psi_{\mu})\ &(0+0)\\
&{\rm gravitino\ multiplet}\        &(\psi_{\mu}, A_\mu)\ & (1+1)\\
&{\rm vector\ multiplet}\        & (A_\mu, \lambda)\ & (1+1)\\
&{\rm chiral\ multiplet}\        & (\varphi,\chi)\        & (1+1)\\
\end{array}
$$
$\varphi$ is real and all fermions are again  real Majorana.

In the massive case 
the little group is $SO(2)$, the same as in the massless case in
$D=4$. Therefore we can adopt the representation theory and assert 
that the massive spin in $D=3$ is the same as helicity in $D=4$.
The only difference is, that one cannot exclude continuous or 
anyonic spin states by topological considerations.
Since half of the supercharges vanish in the massless 
$D=4$ representations \cite{wess} leaving the same number of supercharges
as for massive representations in  $D=3$
we expect that for any $N$ the supermultiplets are the same. 
This will be shown in the following.   

To construct the massive multiplets, it is convenient to introduce the
following linear combinations of the supercharges
 \begin{equation}\label{R}
    R^I:=\frac{1}{\sqrt{2}}(Q_1^I-iQ_2^I)\ , \qquad
   (R^I)^{\dagger}:=\frac{1}{\sqrt{2}}(Q_1^I+iQ_2^I)\ .
  \end{equation}  
These charges have well-defined spin-properties in the sense that
they raise and lower the spin by an amount of $\frac{1}{2}$,
which can be seen as follows. For the single spin operator $S$ in $D=3$
one has by use of $S=M^{12}$ and (\ref{spinor}) 
 \begin{equation}
    [S,R^I] = -\frac{1}{2}R^I  \ ,\qquad
   [S,(R^I)^{\dagger}] = \frac{1}{2}(R^I)^{\dagger},
  \end{equation}
from which we conclude, that for a state $\ket{j}$ with spin $j$, i.e. 
with $S\ket{j} = j\ket{j}$ one has
 \begin{equation}\label{spin1}
   S(R^I)^{\dagger}\ket{j} = (j+\frac{1}{2})(R^I)^{\dagger}\ket{j}\ , \qquad
    SR^I\ket{j} = (j-\frac{1}{2})R^I\ket{j}\ .
   \end{equation}
Therefore the operators $R^I$ and $(R^I)^{\dagger}$ transform bosons
and fermions into each other and anyons into themselves. 

In terms of $R^I$ and $(R^I)^{\dagger}$
the algebra (\ref{susy2}) can be rewritten as
 \begin{equation}\label{RR}
  \begin{split}
   \{R^I,(R^I)^{\dagger}\}&=2P_0 \delta^{IJ}\ , \\
   \{R^I,R^J\}&=-2(P_2-iP_1)\delta^{IJ}\ , \\
   \{(R^I)^{\dagger},(R^J)^{\dagger}\}&=-2(P_2+iP_1)\delta^{IJ}\ .
  \end{split}
 \end{equation}
In the massive case we can boost into the rest frame $P_{\mu}=(m,0,0)$
such that after the rescaling $a^I:=\frac{1}{\sqrt{2m}}R^I$ and
$(a^I)^{\dagger}:=\frac{1}{\sqrt{2m}}(R^I)^{\dagger}$ the algebra
(\ref{RR}) reduces to the well known algebra of fermionic
creation and annihilation operators
 \begin{equation}\label{ferm}
  \{a^I,(a^J)^{\dagger}\}=\delta^{IJ} \ ,\qquad
  \{a^I,a^J\}=\{(a^I)^{\dagger},(a^J)^{\dagger}\}=0\ ,
 \end{equation} 
where again $(a^I)^{\dagger}$ raises the spin by $\frac{1}{2}$ and
$a^I$ lowers the spin by $\frac{1}{2}$. 
Now the construction of the supermultiplets is straightforward. 
We introduce the Clifford vacuum $\Omega$ defined by $a^I\Omega=0$ for 
all $I=1,...,N$  and construct the representation by
application of $(a^I)^{\dagger}$ in the standard fashion \cite{wess}.

For $N=1$ we get only the two linear independent states $\Omega$ and
$a^{\dagger}\Omega$. If $\Omega_j$ has spin $j$, we get a multiplet
with spins $(j,j+\frac{1}{2})$. Similarly for $N=2$ we get multiplets
with spins $(j,j+\frac{1}{2},j+\frac{1}{2},j+1)$.
The multiplet structures are given in tab. \ref{mult1}. 
One sees that the massive $D=3$ multiplets are for arbitrary $N$ 
identical to the massless $D=4$ multiplets, as expected.

\bigskip

\begin{table}[h]
  \begin{center}
   $\begin{array}{c|cccc}
    \text{Spin} & \Omega_0 & \Omega_{1/2} & \Omega_1 & \Omega_{3/2}\\
    \hline
    0 & 1\\
    \frac{1}{2} & 1 & 1\\
    1 & & 1 & 1\\
    \frac{3}{2} & & & 1 & 1\\
    2 & & & & 1
   \end{array}$\hspace{15mm}
   $\begin{array}{c|ccc}
    \text{Spin} & \Omega_0 & \Omega_{1/2} & \Omega_1 \\
    \hline
    0 & 1\\
    \frac{1}{2} & 2 & 1\\
    1 & 1 & 2 & 1\\
    \frac{3}{2} & & 1 & 2 \\
    2 & & & 1 
   \end{array}$
  \end{center}
\caption{Massive Multiplets for $N=1$ and $N=2$}\label{mult1}
 \end{table}

\end{appendix}

\end{document}